# Interpretation of the slight periodic displacement in the Michelson-Morley experiments

Masanori Sato

*Honda Electronics Co., Ltd.,*
*20 Oyamazuka, Oiwa-cho, Toyohashi, Aichi 441-3193, Japan*

E-mail: msato@honda-el.co.jp

Abstract: The slight periodic displacements in the Michelson-Morley interference experiments do not show ether-drift; however, they do show the effects of the theory of general relativity that is, the effects of acceleration and deceleration to the interference condition of the Michelson-Morley interference experiments. Therefore, these slight periodic variations are very significant and important for strengthening the theory of special and general relativity. The slight periodic displacements are discussed from the viewpoint of the theories of general relativity and interference.

1. Introduction

The Michelson-Morley experiment does not measure the speed of light; however, it demonstrates that an interference pattern does not vary according to the motion of the earth. It is clear that a single photon experiment shows the same interference results [1]: only a single photon is in the interferometer, therefore, the light speed difference between the two paths is not detected, as shown in **Fig.1**.

In spite of the fact that the Michelson-Morley experiment is an interference experiment, slight periodic sidereal displacements were detected. Michelson et al. [2] reported that the slight periodic displacements were one fifteenth of 300 km/s. These slight periodic variations were observed by Michelson [2-4], Miller [5], Kennedy [6], and so on. Miller [5], who carefully checked the experimental conditions of thermal effects (artifacts) for measurement, concluded that the ether-drift was observed. Of course, the detected experimental data values were very small as compared to the expected value: in almost all the cases, they were less than 10%. The slight periodic displacements were detected by many reserchers [2-6]. It is important that the displacements change periodically depending on the sidereal time. Miller very carefully checked the thermal effects on the experiments, he paid much attention to heat insulation and carried out a control test using a heater. Miller also discussed the temperature artifacts in a letter with the Editor of Physical Review [7]. However, in 1955, Shankland et al. [8] reported that the slight periodic displacements were caused by thermal artifacts. Thus, the null results have been believed widely, and they became the common belief.

The null results do not agree with the theory of relativity. I do not agree with the null results, because I

consider that the slight periodic displacements are strong evidence, not against the theory of special relativity, but supporting the theory of general relativity.

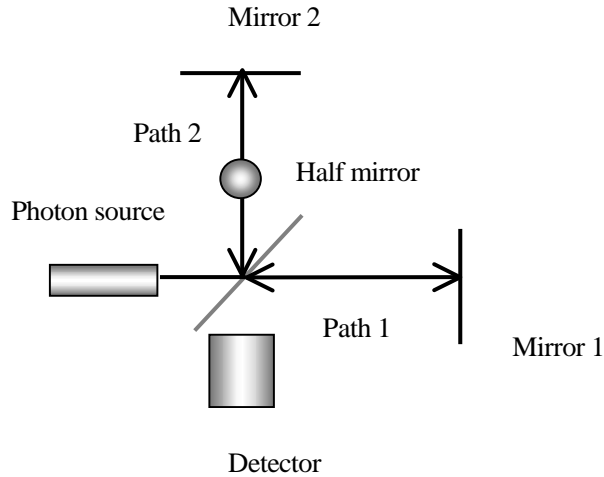

Fig. 1   Conceptual diagram of the Michelson-Morley experiment with single photon [1]. A photon enters the interferometer via the beam splitter, is reflected by the mirror, and is then recombined by the beam splitter. We can detect the interference that is, the photon paths can be arranged such that the detector detects the photons. According to this schematic diagram, a single photon Michelson interferometer appears to detect only the interference, and does not measure the speed of photons. In a single photon interferometer, there is only one photon in the photon paths; therefore, it cannot measure the arrival time of the photons on path 1 and path 2. Only the interference condition is evaluated.

The Global Positioning System (GPS) is used in car navigation systems. The use of special relativity in GPS has been summarized by Ashby [8]. The GPS satellites orbit in a region of low gravity (~20,000 km from ground level) at 4 km/s as shown in **Fig. 2**. In the GPS satellites, either the earth-centered, earth-fixed, reference rame (ECEF frame) or the earth-centered locally inertial (ECI) coordinate system is used for calculations and is operationg well. In the application of the theory of special relativity to GPS satellites, we have to use the ECEF frame, which is practically a reference frame at rest.

Sato [9] pointed out that another reference frame at rest ( for example, one based on the cosmic microwave background) can be applied to GPS satellites experiments on the atomic clock; however, the calculation needs not only special relativity, but also general relativity. This is because the GPS satellites' orbit parallel in **Fig. 2** moves on the cycloid orbits in the the cosmic microwave background shown in **Fig. 3**: the GPS satellite motion is under the periodic sidereal acceleration and deceleration. However, in the GPS satellite experiments on the atomic clock, a large periodic orbital deviation that critically depends on the motion of the orbital plane of the GPS satellites in the cosmic microwave background has not been detected [10]. That is, the difference in gravitational potential between the GPS satellites and the ground causes a 45.7 μs time gain every day, and the effect of the theory of special relativity results in a 7.1 μs time delay every day. However, there is no periodic orbital deviation detected [10].

In this short note, a rough illustration of the interpretation of the slight periodic displacement in the Michelson-Morley experiments is proposed. That is, these slight periodic variations do not show ether-drift and should be discussed from the viewpoint of the theories of general relativity as well as special relativity.

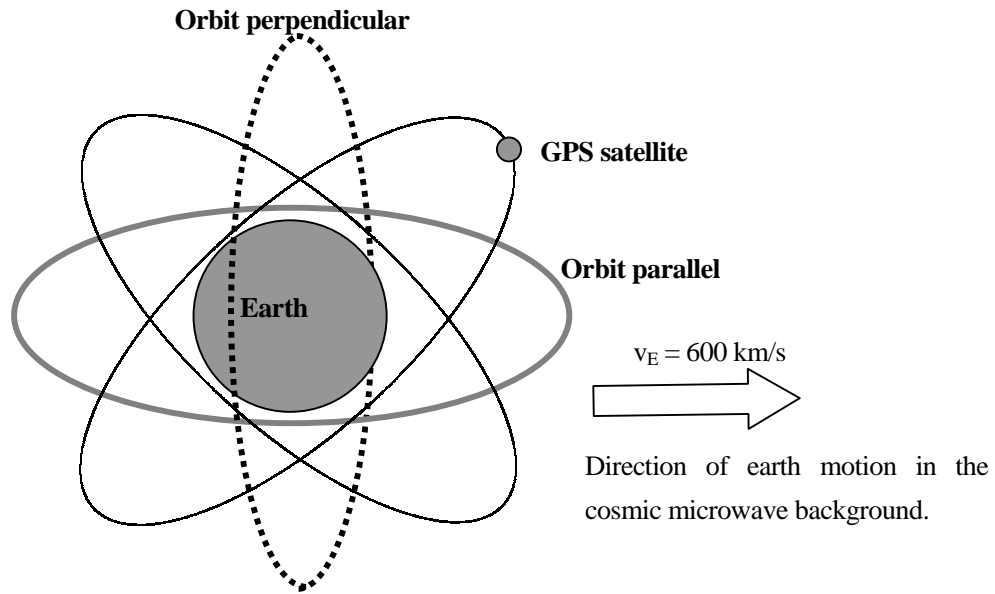

Fig. 2   GPS satellite orbits in the cosmic microwave background [10]
Twenty four satellites are launched on six orbits and the GPS is operating well by the ECEF frame.

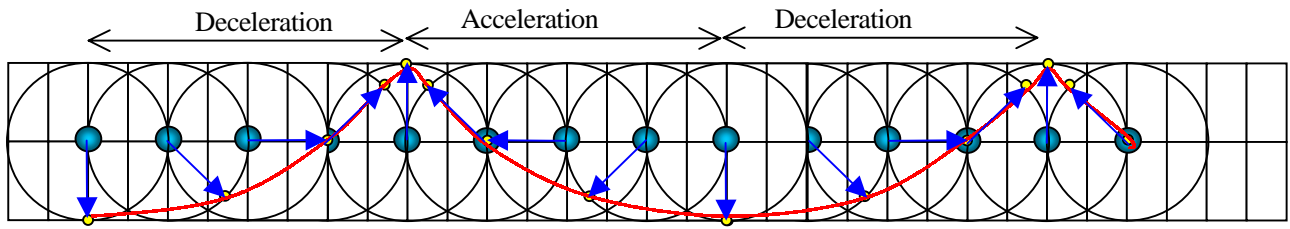

Figure 3   Traveling path of the GPS satellite in an arbitrary reference frame [10].

A clock in the satellite travels according to the red line. From an arbitrary reference frame, the satellite's motion is periodic orbital accelerated motion (that is, cycloid). The reference time of the atomic clock in the GPS satellite does not show any periodic deviation [11]. The Michelson-Morley experimental setup on earth obtains a periodic deviation of the interference condition, which shows the slight periodic displacements in the previous papers.

2. Interpretation of the slight periodic displacement in the Michelson-Morley experiments

The numerical values of the Michelson-Morley experiments were summarized by Shankland [8]: about 10 km/s (Miller [5]), 24 km/s (Kennedy et al. [6]), one-fifteenth of the expected value of 300 km/s (Michelson, Pease, Pearson [3]), and 20 km/s (Michelson, Pease, Pearson [4]). I do not beleive that the Michelson-Morley experiments can detect ether-drift; therefore, I consider these experimental results to show the effects of the accelerated motion of the experimental setups.

As mentioned above, the Michelson-Morley experiment is an interference experiment. A single photon experiment reveals that the interference pattern difference does not show the difference of the light speeds between the two paths. I do not believe that the slight periodic displacement shows ether-drift but the effect of the theory of general relativity.

The speed of light is affected by acceleration according to the theory of general relativity. If the earth moves in the cosmic microwave background, the motion of the experimental setups on earth becomes the cycloid motion shown in **Fig. 4** that is, accelerated motion. The acceleration of cycloid motion critically depends on the earth's rotation in the cosmic microwave background, therefore there is a possibility of sidereal periodic variations of the interference pattern of the of the Michelson-Morley experiment.

The GPS satellites' experiments involving the atomic clock do not show any periodic deviation: the reference time of the atomic clock is not affected by cycloid motion. In the atomic clock experiment, the effects of special relativity (that is, a periodic motion of velocity) and general relativity (the effect of acceleration and deceleration) appear to offset each other. However, in the Michelson-Morley interference experiment, the effects of special relativity and general relativity do not offset each other.

A simple illustration shows that experimental setups on earth are affected by acceleration caused by a cycloid motion: at this stage I cannot illustrate the orbit of the Michelson-Morley experimental setup on Mt. Wilson. The orbit of the experimental setup in the cosmic microwave background will be the orbit of cycloid. At this stage, I do not know how to calculate the effects of the theory of general relativity on the Michelson-Morley experiment.

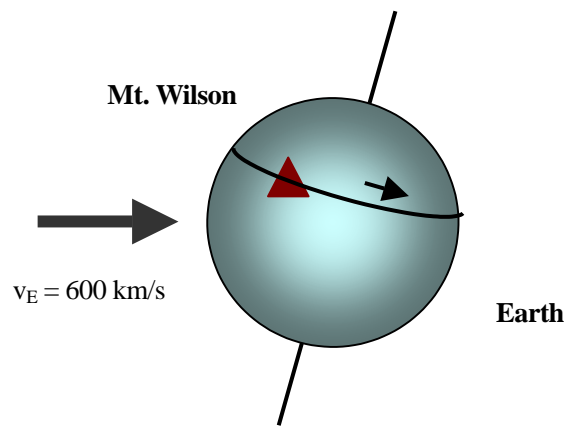

**Fig. 4**  Illustration of the assumption: the earth's motion in the cosmic microwave background. Acceleration is caused by the summation of the drift velocity and the rotation of the earth. The Michelson-Morley experimental setup on Mt. Wilson will detect sidereal periodic acceleration. The earth's motion is assumed to be at some 600 km/s towards the constellation Virgo [11].

The summation of the drift velocity and the rotation of the earth becomes accelerated motion. For example, if the drift velocity $v_E$= 600 km/s [11], and the rotation of the earth is assumed to be as shown in **Fig. 4**, the Michelson-Morley experimental setup on Mt. Wilson will detect sidereal periodic acceleration.

Cycloid motion is the synthesized motion of drift and rotation: the direction of acceleration changes depending on the angle between the drift direction and the rotation axis. For example, if the drift velocity is perpendicular to the rotation axis, the orbit of the earth is helical, and the drift velocity is parallel to the rotation axis, the orbit of the earth is cycloid. In a helical orbit, there is no periodic acceleration detected. In a cycloid orbit, the direction of acceleration is parallel to the drift velocity and periodic acceleration will be detected

3. Conclusion

We pointed out that the slight periodic displacement in the Michelson-Morley experiments is very significant, and should not be dismissed. The slight periodic displacements do not show ether-drift; however, they show the effects of the theory of general relativity that is, the effects of acceleration and deceleration on the interference condition. Therefore, these slight periodic variations are very significant and important for strengthening the theories of special and general relativity. The Michelson-Morley experiments should be discussed from the viewpoint of the theories of general relativity and interference.